%% file: icpv2_resub2.tex
\documentstyle[preprint,eqsecnum,aps,floats,epsfig]{revtex}
\begin{document}
\draft
\input prepri.tex

\title{\boldmath Observation of 
Large $CP$ Violation in the Neutral $B$ Meson System}

\author{
   K.~Abe$^{9}$,               
  K.~Abe$^{37}$,              
  R.~Abe$^{27}$,              
  I.~Adachi$^{9}$,            
  Byoung~Sup~Ahn$^{16}$,      
  H.~Aihara$^{39}$,           
  M.~Akatsu$^{20}$,           
  G.~Alimonti$^{8}$,          
  K.~Asai$^{21}$,             
  M.~Asai$^{10}$
% ...
\\
{\large \bf Belle Collaboration}
}

\address{
% ...
$^{8}${University of Hawaii, Honolulu HI}\\
$^{9}${High Energy Accelerator Research Organization (KEK), Tsukuba}\\
$^{10}${Hiroshima Institute of Technology, Hiroshima}\\
% ...
$^{16}${Korea University, Seoul}\\
% ...
$^{20}${Nagoya University, Nagoya}\\
$^{21}${Nara Women's University, Nara}\\
% ...
$^{27}${Niigata University, Niigata}\\
% ...
$^{37}${Tohoku Gakuin University, Tagajo}\\
% ...
$^{39}${University of Tokyo, Tokyo}\\
% ...
}

\date{\today}
\maketitle
\begin{abstract}
We present a measurement of the
Standard Model $CP$ violation parameter
$\sin 2\phi_1$  based on
a $29.1~{\rm fb}^{-1}$ data sample collected at the $\Upsilon(4S)$ resonance
with the Belle detector at the KEKB asymmetric-energy $e^+e^-$ collider.
One neutral $B$ meson
is fully reconstructed as a
$J/\psi K_S$, $\psi(2S) K_S$, $\chi_{c1} K_S$, $\eta_c K_S$, $J/\psi K_L$
or $J/\psi K^{*0}$
decay  and
the flavor of the accompanying $B$ meson is identified
from its
decay products.
From the asymmetry in the
distribution of the time intervals between the two $B$ meson decay points,
we determine
$\sin 2\phi_1 = 0.99\pm 0.14({\rm stat})\pm 0.06({\rm syst}).$
We conclude that we have observed  
$CP$ violation in the neutral $B$ meson system.
\end{abstract}


\narrowtext

Kobayashi and Maskawa (KM) proposed, in 1973,  a model
where $CP$ violation is
incorporated as an irreducible complex phase in the
weak-interaction quark mixing matrix~\cite{KM}.
The idea, which was presented at a time when only the $u$, $d$ and
$s$ quarks were known to exist,  was remarkable because
it required the existence of six quarks.
The subsequent
discoveries of the $c$, $b$ and $t$ quarks, and the compatibility
of the model with the $CP$ violation observed in the neutral $K$ meson
system led to the incorporation of the KM mechanism into the Standard Model,
even though it had not been conclusively tested
experimentally.

In 1981, Sanda, Bigi and Carter~\cite{carter} pointed out that the KM
model predicted large $CP$ violation in
certain decays of $B$ mesons for a range of quark mixing
parameters.  Subsequent measurements of the $B$ meson lifetime~\cite{MAC}
and the discovery of $B^0\overline{B}{}^0$ mixing~\cite{ARGUS}
indicated that the parameters lie within such a range.
Thus, measurements of $CP$
violation in $B$ meson decays provide important tests of the KM model.

The model predicts
a $CP$ violating asymmetry in the time-dependent
rates for initial $B^0$ and $\overline{B}{}^0$
decays to a common $CP$ eigenstate, $f_{CP}$~\cite{carter}. 
In the case where $f_{CP}=(c\overline{c})K^0$, the asymmetry
is given by
$$
\begin{array}{l}
A(t)\equiv
\frac{\Gamma(\overline{B}{}^0\to f_{CP})-\Gamma({B^0}\to f_{CP})}
{\Gamma(\overline{B}{}^0\to f_{CP})+\Gamma({B^0}\to f_{CP})}
=-\xi_f\sin 2\phi_1 \sin\Delta m_d t,
\end{array}
$$
where $\Gamma(\overline{B}{}^0$ $(B^0)$
$\to f_{CP})$ is the decay rate
for $\overline{B}{}^0(B^0)$ 
to $f_{CP}$ at a proper time $t$ after production,
$\xi_f$ is the $CP$-eigenvalue of $f_{CP}$,
$\Delta m_d$ is the mass difference 
between the two $B^0$ mass eigenstates, and
$\phi_1$  is one of the three internal
angles of the Unitarity Triangle, defined as
$\phi_1\equiv 
\pi-\arg\left(\frac{-V^*_{tb}V_{td}}{-V^*_{cb}V_{cd}}\right)$~\cite{Sanda}.
For the ($c\bar{c})K^0$ decays, both the ambiguity 
due to strong interactions and the contribution from direct $CP$ 
violation are expected to be small~\cite{Sanda}.

Our previous determination,  using a data sample taken in 
1999-2000,
found
$\sin 2\phi_1 =  0.58 ^{+0.32}_{-0.34}{\rm (stat)}$ $ ^{+0.09}_{-0.10} {\rm
(syst)}$~\cite{belle_phi1},
which is consistent with the KM model constraints
from indirect measurements~\cite{SMpred}.
Although the combination of this result with  other
measurements of
$\sin 2\phi_1$~\cite{BaBar} strongly
indicates violation of $CP$ symmetry in $B$ meson decays,
the published results are still not conclusive.
In this Letter we report a new measurement of $\sin 2\phi_1$
that uses improved reconstruction
algorithms and incorporates data taken in 2001 
to achieve a
four-fold increase in the size of the event sample.
The result reported here includes the earlier data
and supersedes the previous value. All data samples have been
analyzed and reconstructed with the same consistent
procedure.

We use a  $29.1~{\rm fb}^{-1}$ data sample,
which contains 31.3 million $B\overline{B}$ pairs, 
collected  with
the Belle detector at the KEKB asymmetric-energy
$e^+e^-$ (3.5 on 8~GeV) collider~\cite{KEKB}.
KEKB operates at the $\Upsilon(4S)$ resonance 
($\sqrt{s}=10.58$~GeV) with
a peak luminosity that exceeds
$4\times 10^{33}~{\rm cm}^{-2}{\rm s}^{-1}$.
The Belle detector is a large-solid-angle magnetic
spectrometer that
consists of a three-layer silicon vertex detector (SVD),
a 50-layer central drift chamber (CDC), a mosaic of
aerogel threshold \v{C}erenkov counters (ACC), time-of-flight
scintillation counters (TOF), and an array of CsI(Tl) crystals
(ECL)  located inside 
a superconducting solenoid coil that provides a 1.5~T
magnetic field.  An iron flux-return located outside of
the coil is instrumented to detect $K_L$ mesons and to identify
muons (KLM).  The detector
is described in detail elsewhere~\cite{Belle}.

We measure $\sin 2\phi_1$
using $B^0\overline{B}{}^0$ meson pairs
produced at the $\Upsilon(4S)$ resonance. 
The two mesons remain in a coherent
$p$-wave state until one of them decays.
The decay of one of the $B$ mesons at time $t_{\rm tag}$
to a final state, $f_{\rm tag}$, which distinguishes between
$B^0$ and $\overline{B}{}^0$, projects the accompanying $B$ meson
onto the opposite $b$-flavor at $t_{\rm tag}$;  this meson
decays to $f_{CP}$ at time $t_{CP}$.
$CP$ violation manifests itself as an asymmetry
$A(\Delta t)$,
where $\Delta t$ is the proper time interval $\Delta t\equiv t_{CP}-t_{\rm tag}$.
At KEKB, the $\Upsilon(4S)$ is produced
with a Lorentz boost of $\beta\gamma=0.425$ nearly along
the electron beamline ($z$).
Since the $B^0$ and $\overline{B}{}^0$ mesons are nearly at 
rest in the $\Upsilon(4S)$ center of mass system (cms),
$\Delta t$ can be determined from the displacement in $z$ 
between the $f_{CP}$ and $f_{\rm tag}$ decay vertices---{\it i.e.}
$\Delta t \simeq (z_{CP} - z_{\rm tag})/\beta\gamma c
 \equiv \Delta z/\beta\gamma c$.

The measurement requires the reconstruction of $B^0\to f_{CP}$
decays, the determination of the $b$-flavor of the accompanying (tagging)
$B$ meson, the measurement of $\Delta t$,
and a fit of the expected $\Delta t$ distribution to 
the measured distribution using a likelihood method.

We reconstruct $B^0$ decays to the following ${CP}$ eigenstates~\cite{CC}:
$J/\psi K_S$, $\psi(2S)K_S$, $\chi_{c1}K_S$, $\eta_c K_S$ for $\xi_f=-1$  and
$J/\psi K_L$ for $\xi_f=+1$.
We also use $B^0\to J/\psi K^{*0}$ decays where
$K^{*0}\to  K_S\pi^0$.  
Here the final state is a mixture of even
and odd $CP$, depending on the relative orbital angular momentum of the
$J/\psi$ and $K^{*0}$.
The $CP$ content is determined from a fit to the full angular
distribution of all $J/\psi K^*$ decay modes other 
than $K^{*0}\to K_S\pi^0$.
We find that the final state is primarily $\xi_f=+1$;
the $\xi_f = -1$ fraction is $0.19 \pm 0.04({\rm stat})\pm 
0.04({\rm syst})$~\cite{Itoh}.

$J/\psi$ and  $\psi(2S)$ mesons are reconstructed via their decays to
$\ell^+\ell^-$ $(\ell=\mu,e)$.
The $\psi(2S)$ is also reconstructed via $J/\psi\pi^+\pi^-$,
and the $\chi_{c1}$ via $J/\psi\gamma$.  The
$\eta_c$ is detected in the $K^+K^-\pi^0$ and
$K_S K^-\pi^+$ modes.
For the $J/\psi K_S$ mode, we use $K_S\to \pi^+\pi^-$ and $\pi^0\pi^0$
decays; for other modes we only use $K_S\to \pi^+\pi^-$.

The $J/\psi$, $\psi(2S)$ and $K_S$ selection has been
described elsewhere~\cite{belle_phi1}.
For $\chi_{c1} K_S$ decays, we select $\chi_{c1}\to  J/\psi\gamma$
decays, rejecting $\gamma$'s that are
consistent with $\pi^0\to \gamma\gamma$
decays, and  use the requirement
$385 < M_{\gamma\ell\ell} - M_{\ell\ell} <430.5~{\rm MeV}/c^2$.
For $\eta_c$ decays, we distinguish kaons from pions
using a combination of CDC energy loss measurements, flight times
measured in the TOF, and the response of the ACC.
Candidate  $\eta_c \rightarrow K^+K^-\pi^0$ $(K_SK^-\pi^+)$ 
decays are selected
with a $KK\pi$ mass requirement that
takes into account the natural width of the $\eta_c$.
For $J/\psi K^{*0}(K_S\pi^0)$ decays, we use $K_S\pi^0$ combinations 
that have an invariant mass 
within 75~MeV/$c^2$ of the nominal $K^*$ mass.
We reduce background from low-momentum $\pi^0$'s by
requiring $\cos\theta_{K^*}<0.8$, where $\theta_{K^*}$ is 
the angle between the $K_S$  momentum vector
and the $K^{*0}$ flight direction calculated  in the $K^{*0}$ rest frame.

We identify $B$
decays using
the energy difference $\Delta E\equiv E_B^{\rm cms} - E_{\rm beam}^{\rm cms}$
and the beam-energy constrained
mass $M_{\rm bc}\equiv\sqrt{(E_{\rm beam}^{\rm cms})^2-(p_B^{\rm cms})^2}$,
where $E_{\rm beam}^{\rm cms}$ is the cms beam energy,
and $E_B^{\rm cms}$ and $p_B^{\rm cms}$ are the cms energy and momentum
of the $B$ candidate.

\begin{figure}
\begin{center}
\epsfxsize 2.8 truein \epsfbox{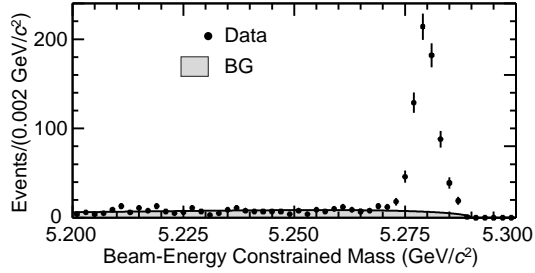}
\end{center}
\caption{The beam-energy constrained mass distribution for
all decay modes combined other than $J/\psi K_L$.
The shaded area is the estimated background.
The signal region is the range $5.27-5.29$ GeV/$\rm{c}^2$.}
\label{fig:bmass}
\end{figure}

Figure~\ref{fig:bmass} shows the combined $M_{\rm bc}$
distribution for
all channels other than $J/\psi K_L$ after
a mode-dependent requirement on  $\Delta E$.
The $B$ meson signal region is defined as
$5.270<M_{\rm bc}<5.290~{\rm GeV}/c^2$.
Table~\ref{tab:tally} lists
the numbers of  observed candidates ($N_{\rm ev}$) and
the background ($N_{\rm bkgd}$) determined by
extrapolating the rate in the
non-signal  $\Delta E$ {\em vs.} $M_{\rm bc}$ region
into the signal region.

Candidate $B^0\to J/\psi K_L$  decays are selected by requiring
ECL and/or KLM hit patterns that are consistent with the 
presence of
a shower induced by a neutral hadron.
The centroid of the shower is required to be in a 45$^\circ$ cone
centered on the $K_L$ direction that is inferred from two-body decay
kinematics and the measured four-momentum of the $J/\psi$.
We reduce the background by means of a likelihood ratio
that depends on the $J/\psi$ cms momentum,
the angle between the $K_L$ and its nearest-neighbor charged track,
the charged track multiplicity of the event, the extent to which the event is
consistent with a $B^+ \to$ $J/\psi K^{*+}(K_L\pi^+)$ hypothesis,
and the polar angle with respect to the $z$ direction
of the reconstructed $B^0$ meson in the cms.
In addition, we remove events that are reconstructed as
$B^0 \to J/\psi K_S$, $J/\psi K^{*0}(K^+\pi^-, K_S \pi^0)$,
$B^+\to$ $J/\psi K^+$, or  $J/\psi K^{*+}(K^+ \pi^0$, $K_S \pi^+)$
decays.  Finally, $K_L$ clusters with positions that match photons
from reconstructed $\pi^0$'s are also rejected.

\begin{figure}
\begin{center}
\epsfxsize 2.8 truein \epsfbox{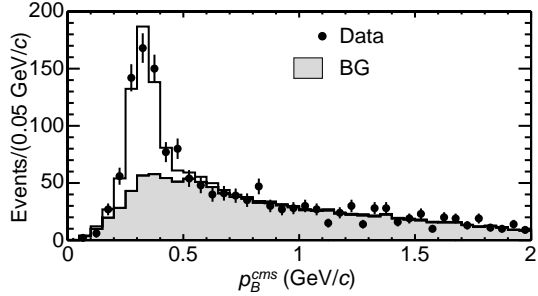}
\end{center}
\caption{The $p_B^{\rm cms}$ distribution for $B^0\to J/\psi K_L$ candidates
with the results of the fit.
The solid line is the signal plus background;
the shaded area is background only.
The signal region for KLM (ECL-only) clusters
is $0.2\leq p_B^{\rm cms}\leq 0.45 (0.40)~~{\rm GeV}/c$.}
\label{fig:pbstar}
\end{figure}

Figure~\ref{fig:pbstar} shows the $p_B^{\rm cms}$ distribution,
calculated with the $B^0 \to J/\psi K_L$ two-body decay hypothesis.
The histograms are the  results of a fit to the signal
and background distributions.   The shapes are
derived from Monte Carlo simulations (MC)~\cite{MC},
and  the normalization and peak position of the signal 
are allowed to vary.
There are 397 entries in the
$0.2\leq p_B^{\rm cms}\leq 0.45~~{\rm GeV}/c$
signal region with KLM clusters. 
There are 172 entries in the 
range $0.2\leq p_B^{\rm cms}\leq 0.40~~{\rm
GeV}/c$ with clusters in the ECL only.
The fit finds a total of $346\pm 29$~$J/\psi K_L$ signal events,
and a signal purity of 61\%.

Leptons, charged pions, and kaons
that are not associated with a reconstructed
$CP$ eigenstate decay are used to identify
the flavor of the accompanying $B$ meson.
Initially,
the $b$-flavor determination is performed at the track level.
Several categories of well measured tracks
that distinguish the $b$-flavor by the track's charge are selected:
high momentum leptons
from  $b\to$ $c\ell^-\overline{\nu}$,
lower momentum leptons from  $c\to$ $s\ell^+\nu$,
charged kaons and $\Lambda$ baryons from $b\to$ $c\to$ $s$,
high momentum pions that originate from decays of the type
$B^0\to$ $D^{(*)-}(\pi^+, \rho^+$, $a_1^+, {\rm etc.})$, and
slow pions from $D^{*-}\to$ $\overline{D}{}^0\pi^-$.
We use the MC to determine a category-dependent variable
that indicates whether
a track originates from a $B^0$ or $\overline{B}{}^0$.
The values of this variable range from 
$-1$ for a reliably identified $\overline{B}{}^0$
to $+1$ for a reliably identified $B^0$ and  depend
on the tagging particle's  charge,
cms momentum, polar angle and
particle-identification probability, as well as other kinematic and
event shape quantities. 
The results from the separate
track categories are then combined to 
take into account correlations
in the case of multiple track-level tags.
This stage determines two event-level parameters, $q$ and $r$.
The first, $q$, has the discrete values $q = +1$  when the tag-side $B$ meson
is more likely to be a $B^0$ and $-1$ when it is more likely to be a
$\overline{B}{}^0$.
The parameter $r$ is an event-by-event flavor-tagging dilution factor which
ranges from $r=0$ for no flavor discrimination to $r=1$ for
unambiguous flavor assignment.
It is used only to sort data into six intervals of $r$, according to 
flavor purity;
the wrong-tag probabilities for the final fit are determined from data.

The probabilities of an
incorrect flavor assignment, $w_l\ (l=1,6)$,
are determined directly from the data for the six  $r$ intervals
using exclusively reconstructed, self-tagged
$B^0\to D^{*-}\ell^+\nu$, $D^{(*)-}\pi^+$,  
$D^{*-}\rho^+$  and $J/\psi K^{*0}(K^+\pi^-)$ decays.
The $b$-flavor of the accompanying $B$ meson
is assigned according to the flavor-tagging algorithm described above.
The exclusive decay and tag vertices are reconstructed
using the same vertexing algorithm that is used in
the analysis to measure $CP$ asymmetry.
The values of
$w_l$ are obtained from the amplitudes of the
time-dependent $B^0\overline{B}{}^0$ mixing oscillations:
$(N_{\rm OF} - N_{\rm SF})/(N_{\rm OF}+N_{\rm SF})
=(1-2w_l )\cos (\Delta m_d \Delta t)$.
Here $N_{\rm OF}$ and $N_{\rm SF}$ are the numbers of opposite and same
flavor events.
We fix $\Delta m_d$ at the world average value~\cite{PDG}.
Table~\ref{tab:tag} lists the resulting  $w_l$ values
together with the fraction of the events ($f_l$)
in each $r$ interval.
The total effective tagging efficiency is
$\sum_l f_l(1-2w_l)^2 = 
0.270\pm 0.008\rm{(stat)}^{+0.006}_{-0.009}\rm{(syst)}$.

The vertex positions for the $f_{CP}$ and $f_{\rm tag}$ decays are
reconstructed using tracks
that have at least one
three-dimensional coordinate determined from 
associated $r$-$\phi$ and $z$
hits in the same SVD layer
along with one or more additional $z$ hits in the other layers.
Each vertex position is required to be
consistent with the interaction point profile smeared in the
$r$-$\phi$ plane by the $B$ meson decay length.
The $f_{CP}$ vertex is determined using
lepton tracks  from
$J/\psi$ or $\psi(2S)$ decays, or prompt tracks from $\eta_c$ decays.
The $f_{\rm tag}$ vertex
is determined from well reconstructed tracks not assigned to $f_{CP}$.
Tracks that form a $K_S$ are not used.
The MC indicates that the typical vertex-finding efficiency and
vertex resolution (rms) for $z_{CP}$ ($z_{\rm tag}$) are
$92~(91)\%$ and $75~(140)~\mu{\rm m}$, respectively.

The proper-time interval resolution for the signal, $R_{sig}(\Delta t)$, 
is obtained by convolving a sum of
two  Gaussians (a {\it main} component
due to  the SVD vertex resolution and charmed meson lifetimes,
plus a {\it tail} component caused by poorly reconstructed tracks)
with a function that takes into account  
the cms motion of the $B$ mesons.
The fraction in the main Gaussian
is determined to be $0.97\pm 0.02$ from a
study of $B^0\to D^{*-}\pi^+$, $D^{*-}\rho^+$,  $D^-\pi^+$,
$J/\psi K^{*0}$, $J/\psi K_S$ and 
$B^+\to \overline{D}{}^0\pi^+$,  $J/\psi K^+$ events.
The means
($\mu_{\rm main}$, $\mu_{\rm tail}$)
and widths
($\sigma_{\rm main}$, $\sigma_{\rm tail}$)
of the Gaussians are
calculated event-by-event from the
$f_{CP}$ and $f_{\rm tag}$ vertex fit error matrices
and the $\chi^2$ values of the fit;
typical values are $\mu_{\rm main}=-0.24~{\rm ps}$,
$\mu_{\rm tail}=0.18~{\rm ps}$
and
$\sigma_{\rm main}=1.49~{\rm ps}$, $\sigma_{\rm tail}=3.85~{\rm ps}$.
The background resolution $R_{bkg}(\Delta t)$ has the same functional
form but the parameters are obtained from a sideband region
in $M_{bc}$ and $\Delta E$.
We obtain  lifetimes for the neutral and charged $B$ mesons 
using the same procedure;
the results~\cite{Deltam} agree well with the world average values.

\begin{figure}
\begin{center}
\epsfxsize 3.1 truein \epsfbox{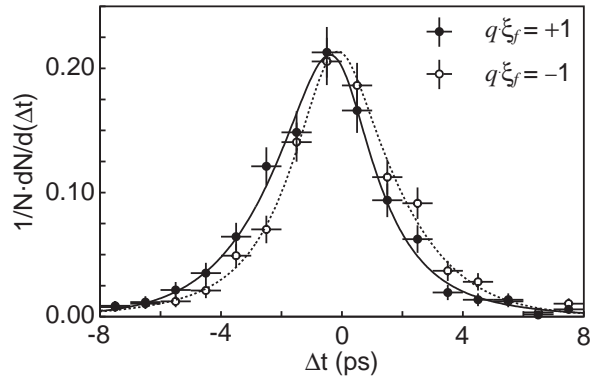}
\end{center}
\caption{
$\Delta t$ distributions 
for the events with $q\xi_f = +1$ (solid
points) and $q\xi_f = -1$ (open points). The 
results of the global fit (with  $\sin 2\phi_1 = 0.99$)
are shown as solid and dashed curves, respectively.
}
\label{fig:dNdt}
\end{figure}

After vertexing we find  560 events with $q=+1$ flavor tags  and 
577 events with $q=-1$. 
Figure~\ref{fig:dNdt} shows the  
observed $\Delta t$ distributions
for the $q\xi_f =+1$ 
(solid points) and 
$q\xi_f = -1$ (open points) event samples.
There is a clear asymmetry between the two distributions; this 
demonstrates that $CP$ symmetry is violated.

We determine 
$\sin 2\phi_1$ 
by performing an
unbinned maximum-likelihood fit of a $CP$ violating
probability density function (pdf) to the observed $\Delta t$ distributions.
For modes other than $J/\psi K^{*0}$
the pdf expected for the signal is
$$
\begin{array}{l}
{\cal P}_{\rm sig}(\Delta t,q,w_l,\xi_f)\\
\quad=\frac{e^{-|\Delta t|/\tau_{B^0}}}{2\tau_{B^0}}
\{1-\xi_f q(1-2w_l)\sin 2\phi_1\sin (\Delta m_d\Delta t )\},
\end{array}
$$
where we fix $\tau_{B^0}$ and $\Delta m_d$ at  their world average
values~\cite{PDG}.
The pdf used for the background distribution is
${\cal P}_{\rm bkg}(\Delta t)
=f_\tau e^{-|\Delta t|/\tau_{\rm bkg}}/2
\tau_{\rm bkg}+(1-f_\tau)\delta(\Delta t),$
where $f_\tau$ is the fraction of the background component
with an effective lifetime $\tau_{\rm bkg}$ and $\delta$ is the Dirac delta
function.
For all $f_{CP}$ modes other than $J/\psi K_L$,
a study using events in background-dominated
regions of $\Delta E$ {\em vs.}  $M_{\rm bc}$ shows
that $f_{\tau}$ is negligibly small.
For these modes,
${\cal P}_{\rm bkg}(\Delta t) = \delta(\Delta t)$.

The $J/\psi K_L$  background is dominated 
by $B \to J/\psi X$ decays where
some final states are $CP$ eigenstates.
We estimate  the fractions of the background  components with and
without
a true $K_L$ cluster by fitting  the $p_B^{\rm cms}$ distribution 
to the expected
shapes determined from the MC.   We also use the MC
to determine the fraction of events with 
definite $CP$ content within each
component.

The result is a background that is 71\%  non-$CP$ modes
with $\tau_{\rm bkg}=\tau_B$.
For the $CP$-mode backgrounds we use the signal pdf given 
above with the appropriate $\xi_f$ values. 
For $J/\psi K^*(K_L\pi^0)$, which is $13\%$ of the background,
we use the $\xi_f=-1$ content determined from 
the full $J/\psi K^*$ sample.
The remaining backgrounds are
$\xi_f=-1$ states ($10\%$) including  $J/\psi K_S$,  
and   $\xi_f=+1$ states  ($5\%$) including 
$\psi(2S)K_L$,  $\chi_{c1}K_L$ and $J/\psi \pi^0$.

For the $J/\psi K^*$ mode, we include
 the  $\Delta t$ and
transversity angle $\theta_{\rm tr}$~\cite{transversity} 
distributions in the likelihood~\cite{Itoh}.  
We use the $\xi_f$ content 
determined from the full angular analysis.

Each pdf is convolved with the appropriate $R(\Delta t)$ to determine
the likelihood value for each event as a function of $\sin 2\phi_1$:
$$
\begin{array}{lcl}
P_i&=&{\displaystyle
\int}\{f_{\rm sig}{\cal P}_{\rm sig}(\Delta t^\prime,
q,w_l,\xi_f) R_{\rm sig}(\Delta t-\Delta t^\prime) \\
&&+(1-f_{\rm sig}){\cal P}_{\rm bkg}(\Delta t^\prime)
R_{\rm bkg}(\Delta t-\Delta t^\prime)\} 
d\Delta t^\prime,
\end{array}
$$
where $f_{\rm sig}$ is the probability that the event is signal,
calculated as a function of $p_B^{\rm cms}$ for $J/\psi K_L$ and
of $\Delta E$ and $M_{\rm bc}$ for other modes.
The only free parameter is $\sin 2\phi_1$, which is
determined by maximizing
the likelihood function
$L=\prod_i P_i$, where the product is over all
events. 

\begin{figure}
\begin{center}
\epsfxsize 2.8 truein \epsfbox{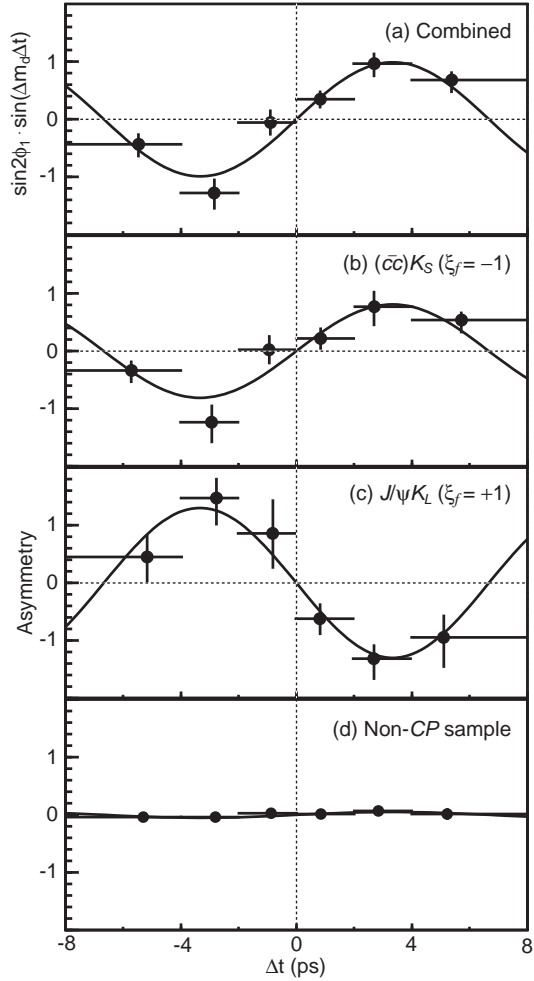}
\end{center}
\caption{(a) The asymmetry obtained
from separate fits to each $\Delta t$ bin for 
the full data sample; the curve is the result of 
the global fit. The
corresponding plots for the (b) $(c\bar{c})K_S$ ($\xi_f=-1$), (c) 
$J/\psi K_L$ ($\xi_f = +1$), and (d) $B^0$ control samples
are also shown.  The curves
are the results of the fit applied separately to the
individual data samples.
}
\label{fig:asym}
\end{figure}

The result of the fit is 
$$  \sin 2\phi_1 = 
0.99 \pm 0.14 ({\rm stat}) \pm 0.06({\rm syst}).$$
In Fig.~\ref{fig:asym}(a) we show the 
asymmetries for the combined data sample 
that are obtained by applying the fit to the events in each
$\Delta t$ bin separately.  The smooth curve is the result
of the global unbinned fit.
Figures~\ref{fig:asym}(b) and (c) show the 
corresponding asymmetries
for the $ (c\bar{c})K_S$  ($\xi_f=-1$)  
and the $J/\psi K_L$ ($\xi_f=+1$) modes separately. 
The observed
asymmetries for the different $CP$ states are opposite,
as expected.
The curves are the results of unbinned fits
applied separately to the two samples; the resultant
$\sin 2\phi_1$ values are
$0.84 \pm 0.17$(stat) and $1.31 \pm 0.23 $(stat), respectively.

The systematic error is dominated by
uncertainties due to effects of the tails of the vertex 
distributions, which contribute $0.04$.
Other significant contributions come from uncertainties (a) in $w_l~(0.03)$;
(b) in the resolution function parameters ($0.02$); and
(c) in the $J/\psi K_L$ background fraction ($0.02$).
The errors introduced by uncertainties in 
$\Delta m_d$ and 
$\tau_{B^0}$ are $0.01$ or less.

We performed a number of checks on the measurement.  
Table~\ref{tab:checks} lists the results obtained by applying the
same analysis to various subsamples.
All values are statistically consistent with each other.
The result is unchanged if we use the $w_l$'s determined
separately for $f_{\rm tag}=B^0$ and $\overline{B}{}^0$.
Fitting
to the non-$CP$ eigenstate self-tagged modes 
$B^0\to D^{(*)-}\pi^+$, $D^{*-}\rho^+$, $J/\psi 
K^{*0}(K^+\pi^-)$ and $D^{*-}\ell^+\nu$,
where no asymmetry is expected,
yields $0.05 \pm 0.04$.  The asymmetry distribution for
this control sample is shown in Fig.~~\ref{fig:asym}(d).
As a further check, we used three independent $CP$ fitting programs and
two different algorithms for the $f_{\rm tag}$ vertexing and found
no discrepancy.

We conclude that there is large $CP$
violation in the neutral $B$ meson system.  
A zero value for
$\sin 2\phi_1$ is ruled out at a level greater than $6\sigma$.
Our result is consistent
with the higher range of values allowed
by the constraints of the KM model as well as with our previous
measurement.

We wish to thank the KEKB accelerator group
for the excellent
operation of the KEKB accelerator.
We acknowledge support from the Ministry of Education, Culture, Sports, Science and
Technology of Japan and
the Japan Society for the Promotion of Science;
the Australian Research Council and the Australian Department of Industry,
Science and Resources;
the Department of Science and Technology of India;
the BK21 program of the Ministry of Education of Korea and
the Center for High Energy Physics sponsored by the KOSEF;
the Polish State Committee for Scientific Research
under contract No.2P03B 17017;
the Ministry of Science and Technology of Russian Federation;
the National Science Council and the Ministry of Education of Taiwan;
and the U.S. Department of Energy.


\newpage
\begin{table}
\caption{The numbers of  observed
events ($N_{\rm ev}$) and the estimated 
background ($N_{\rm bkgd}$)
in the signal region for each $f_{CP}$ mode.}
\label{tab:tally}
\begin{tabular}{lrr}
Mode & $N_{\rm ev}$ & $N_{\rm bkgd}$\\
\hline
$J/\psi(\ell^+\ell^-) K_S(\pi^+\pi^-)$ & 457 & 11.9\\
$J/\psi(\ell^+\ell^-) K_S(\pi^0\pi^0)$  & 76 & 9.4\\
$\psi(2S)(\ell^+\ell^-)K_S(\pi^+\pi^-)$  & 39 & 1.2\\
$\psi(2S)(J/\psi\pi^+\pi^-)K_S(\pi^+\pi^-)$ & 46 & 2.1\\
$\chi_{c1}(J/\psi\gamma) K_S(\pi^+\pi^-)$ & 24 & 2.4\\
$\eta_c(K^+K^-\pi^0)K_S(\pi^+\pi^-)$ & 23 & 11.3\\
$\eta_c(K_S K^-\pi^+)K_S(\pi^+\pi^-)$ &41 & 13.6\\
$J/\psi K^{*0}(K_S\pi^0)$& 41 & 6.7\\
\hline
Sub-total & 747 & 58.6	\\
\hline
$J/\psi(\ell^+\ell^-) K_L$ & 569 & 223
\end{tabular}
\end{table}

\begin{table}
\caption{
The event fractions ($f_l$)
and incorrect flavor assignment probabilities ($w_l$)
for each $r$ interval. The errors include both statistical
and systematic uncertainties.}
\label{tab:tag}
\begin{tabular}{lccc}
$l$&$r$ & $f_l$ & $w_l$\\
\hline
1&$0.000-0.250$ & $0.405$ & $0.465^{+0.010}_{-0.009}$ \\
2&$0.250-0.500$ & $0.149$ & $0.352^{+0.015}_{-0.014}$ \\
3&$0.500-0.625$ & $0.081$ & $0.243^{+0.021}_{-0.030}$\\
4&$0.625-0.750$ & $0.099$ & $0.176^{+0.022}_{-0.017}$\\
5&$0.750-0.875$ & $0.123$ & $0.110^{+0.022}_{-0.014}$\\
6&$0.875-1.000$ & $0.140$ & $0.041^{+0.011}_{-0.010}$
\end{tabular}
\end{table}

\begin{table}
\caption{
The values of
$\sin 2\phi_1$  for various  subsamples
(statistical errors only).
}
\label{tab:checks}
\begin{tabular}{ll}
Sample & $\sin 2\phi_1$\\
\hline
$f_{\rm tag}=B^0$ ($q=+1$) &  $0.84\pm 0.21$\\
$f_{\rm tag}=\overline{B}{}^0$ ($q=-1$) & $1.11\pm 0.17$\\
\hline
$J/\psi K_S(\pi^+\pi^-)$ & $0.81\pm 0.20$\\
$(c\bar{c})K_S$ except $J/\psi K_S(\pi^+\pi^-)$ & $1.00 \pm 0.40$\\
$J/\psi K_L$  & $1.31\pm 0.23$\\
$J/\psi K^{*0}(K_S\pi^0)$ & $0.85 \pm 1.45$\\
\hline 
All & $0.99\pm 0.14$
\end{tabular}
\end{table}

%
%

%
%

\end{document}

%% file: prepri.tex
%
%
%
%
\topmargin -0.5in
\makeatletter
\def\maketitle{\par
\begingroup
\let\cite\@bylinecite
\def\thefootnote{\fnsymbol{footnote}}%
\if@twocolumn
\twocolumn[\@maketitle\vskip2pc]%
\else
\newpage
\epsfysize3cm
\epsfbox{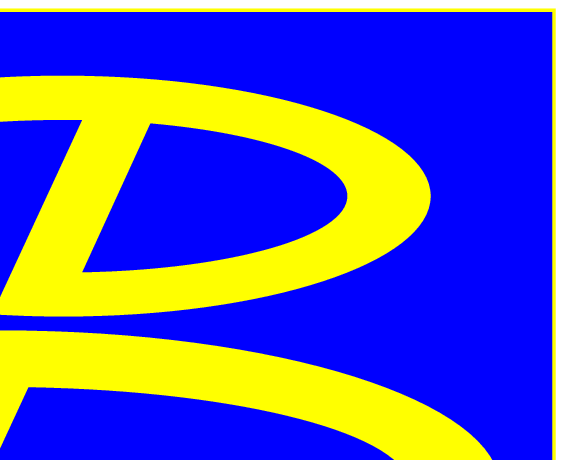}    
\global\@topnum\z@ %
\@maketitle
\fi
\thispagestyle{plain}\@thanks
\endgroup
\def\thefootnote{\arabic{footnote}}%
\setcounter{footnote}{0}%
\let\maketitle\relax \let\@maketitle\relax
\let\@thanks\relax \let\@authoraddress\relax \let\@title\relax
\let\@date\relax \let\thanks\relax
}
\makeatother

\tighten

\draft 


\preprint{\tighten\vbox{\hbox{\hfil KEK preprint 2001-50}
                        \hbox{\hfil Belle preprint 2001-10}
                        \hbox{\hfil }
                        \hbox{\hfil }\hbox{\hfil }\hbox{\hfil }
}}

%% file: icpv2_resub2.bbl
\begin{references}

\bibitem{KM}
M.~Kobayashi and T.~Maskawa, Prog. Theor. Phys. {\bf 49}, 652 (1973).

\bibitem{carter}
A.B.~Carter and A.I.~Sanda, Phys. Rev. {\bf D23}, 1567 (1981);
I.I.~Bigi and A.I.~Sanda, Nucl. Phys. {\bf B193}, 85 (1981).

\bibitem{MAC}
E.~Fernandez {\it et al.} (MAC Collab.), Phys. Rev. Lett. {\bf 51},
1022 (1983); N.~Lockyer {\it et al.} (Mark II Collab.),
Phys. Rev. Lett. {\bf 51}, 1316 (1983).

\bibitem{ARGUS} H.~Albrecht {\it et al.} (ARGUS Collab.),
Phys. Lett. {\bf B192}, 245 (1987).

\bibitem{Sanda}
H.~Quinn and A.I.~Sanda, Eur. Phys. Jour. {\bf C15}, 626 (2000).
(This angle is also known as $\beta$.)

\bibitem{belle_phi1}
A.~Abashian {\it et al.} (Belle Collab.),  Phys. Rev. Lett. {\bf
86}, 2509 (2001).

\bibitem{SMpred}
For example: M.~Ciuchini {\it et al.}, hep-ph/0012308, submitted
for publication to JHEP.

\bibitem{BaBar}
B.~Aubert {\it et al.} (BaBar Collab.),  Phys. Rev. Lett. {\bf 86}, 2515
(2001) reports $\sin 2\phi_1 = 0.34 \pm 0.20 \pm 0.05$;
T.~Affolder {\it et al.} (CDF Collab.), Phys. Rev. {\bf D61},
072005 (2000) reports $\sin 2\phi_1 = 0.79^{+ 0.41}_{-0.44}$;
R.~Barate {\it et al.} (ALEPH Collab.), Phys. Lett. {\bf B492}, 259
(2000) reports $\sin 2\phi_1 = 0.84^{+0.82}_{-1.04}\pm 0.16$; and
K.~Ackerstaff {\it et al.} (OPAL Collab.), Eur. Phys. Jour. {\bf C5}, 379
(1998) reports $\sin 2\phi_1 = 3.2^{+1.8}_{-2.0} \pm 0.5$.

\bibitem{KEKB}
KEKB B Factory Design Report, KEK Report 95-1, 1995, unpublished.

\bibitem{Belle}
K.~Abe {\it et al.} (Belle Collab.),
{\em The Belle Detector}, KEK Report 2000-4, to be published
in Nucl. Instrum. Methods.

\bibitem{CC}
Throughout this Letter, whenever a mode is quoted the inclusion 
of the charge conjugate mode is implied.


\bibitem{Itoh}
K.~Abe {\it et al.} (Belle Collab.),
{\em Measurements of Polarization and $CP$ Asymmetry
in $B\to J/\psi + K^*$ decays}, paper submitted to LP01,
Rome, July 2001; BELLE-CONF-0105.


\bibitem{MC}
We use the QQ $B$ meson decay event
generator developed~by~the~CLEO~Collaboration

\noindent
(http://www.lns.cornell.edu/public/CLEO/soft/QQ)

\noindent
and GEANT3 for the detector simulation;
CERN Program Library Long Writeup  W5013, CERN, 1993.

\bibitem{PDG}
D.E. Groom {\it et al.} (Particle Data Group), 
Eur. Phys. J. {\bf C15}, 1 (2000).

\bibitem{Deltam} The measured
$B$-lifetimes are: $\tau_{B^0} = 1.547 \pm 0.021$~ps and 
$\tau_{B^+} = 1.641 \pm 0.033$~ps (statistical errors only). 

\bibitem{transversity}
$\theta_{\rm tr}$ is defined as the angle 
between the $\ell^+$ direction in the $J/\psi$ rest frame and the 
$z$-axis,
where the $x$-axis is defined as the direction of motion of the
$J/\psi$ in the $\Upsilon(4S)$
rest frame. The $x$-$y$ plane is defined by the $K^*$ decay
products  in the $J/\psi$ rest frame.


\end{references}
